\documentclass[12pt]{iopart}

\usepackage{iopams}  
\usepackage{amssymb}
\usepackage{amsthm}
\usepackage[dvipdfmx]{graphicx}
\usepackage{hyperref}
\usepackage{dcolumn}
\usepackage{bm}
\usepackage{multirow}
\usepackage{comment}
\usepackage{array,booktabs}
\usepackage{braket}
\usepackage{comment}
\usepackage{cite}
\usepackage{color}
\usepackage{xcolor}
\usepackage{url}
\hypersetup{
    colorlinks=true,
    citecolor=blue,
    linkcolor=blue
}

\begin{document}

\title[The cavity method to protein design problem]{
The cavity method to protein design problem}

\author{Tomoei Takahashi $^1$, George Chikenji $^2$, and Kei Tokita $^1$}


\address{$^1$Graduate School of Informatics, Nagoya University, Nagoya, 464-8601, Japan.}
\address{$^2$Graduate School of Engineering, Nagoya University, Nagoya, 464-8603, Japan.}


\ead{takahashi@phys.cs.i.nagoya-u.ac.jp}
\address{DOI : \url{https://doi.org/10.1088/1742-5468/ac9465}}

\vspace{10pt}
\begin{indented}
\item[]
\end{indented}

\begin{abstract}
In this study, we propose an analytic statistical mechanics approach to solve a fundamental problem in biological physics called protein design.  Protein design is an inverse problem of protein structure prediction, and its solution is the amino acid sequence that best stabilizes a given conformation.  Despite recent rapid progress in protein design using deep learning, the challenge of exploring protein design principles remains.  Contrary to previous computational physics studies, we used the cavity method, an extension of the mean-field approximation that becomes rigorous when the interaction network is a tree.   We found that for small two-dimensional (2D) lattice hydrophobic-polar (HP) protein models, the design by the cavity method yields results almost equivalent to those from the Markov chain Monte Carlo method with lower computational cost.
\end{abstract}

%
\vspace{2pc}

\tableofcontents
\addtocontents{toc}{\protect\thispagestyle{empty}}
\pagestyle{empty}

\section{Introduction}

Over the past 25 years, there has been an increasing interest in using statistical mechanics approaches to solve inverse problems in information processing, such as error-correcting codes, combinatorial optimization problems, data analysis, and machine learning \cite{mezard2009information}.  The Ising model or a spin glass model, which is the model of disordered systems with heterogeneous interactions, has been commonly used to solve inverse problems \cite{mezard1987spin-glasses}.  The calculation is performed in the opposite direction of the ordinary statistical mechanics, i.e., to identify a microscopic sequence or parameters that minimize specific cost functions.  For example, the Boltzmann machine learning identifies synaptic weights which makes learning patterns an equilibrium state; in an inverse Ising problem we infer the coupling strengths between spins given observed spin configurations \cite{chau2017inverse}.  An essential benefit of these studies is that such approaches have solved the problem of computational explosion in probabilistic inference for information processing using approximation methods, such as mean-field approximation.  

In this paper, we apply statistical mechanics to an inverse problem, not to an information processing problem, but to a fundamental problem in biological phenomena: protein design.  Protein design is an inverse problem of protein structure prediction.  The problem is to find the amino acid sequence that best stabilizes a given conformation \cite{coluzza2017computational, cocco2018inverse}.

The application of protein design to drug design by designing novel proteins with desired biological functions is significant.  Besides, its fundamental scientific significance is understanding the relationship between structures and sequences from the perspective of ``design" which cannot be clarified by the forward approach, i.e., structure prediction. 

In recent years, there has been much research and success in protein design through deep learning \cite{ovchinnikov2021structure}.  Despite this, the fundamental questions of the principle of protein design or how protein sequences have evolved remain essential.  The main interest of this research is to address these fundamental questions while proposing specific methods and designing model proteins.  Such principle studies are significant even for analysis using vast amounts of data, including deep learning.

Because a native conformational state of a protein is an equilibrium state determined solely by the amino acid sequence under the physiological conditions (the Anfinsen’s dogma) \cite{anfinsen1973principles}, to solve the protein design problem, it is suﬀicient to know the sequence that minimizes the free energy of a given native conformation.  Therefore, the solution to the protein design problem is a sequence that makes the given native conformation the only ground state at low temperature.  

The ``design criteria” can be categorized into two main ways: maximizing target probability (MTP) and energy minimization.  The lattice HP model \cite{lau1989lattice}, a kind of Ising model of proteins is the most commonly used.  The MTP is a method to maximize the conditional probability of a native conformation expressed by a canonical distribution (target probability) \cite{kurosky1995design,deutsch1996new}.  Moreover, various MTP-based methods have been proposed \cite{seno1996optimal,irback1998monte, irback1999design, iba1998design, tokita2000dynamical}.  Energy minimization is a more straightforward method and a sequence that minimizes the energy of a given conformation is the solution  \cite{shakhnovich1993engineering,salvi2002design, abeln2008disordered, ni2013interplay, abeln2014simple, bianco2017role, bianco2019protein, bianco2020in}.  The aforementioned studies  are engineering-oriented and computational approaches.

By contrast, we propose an analytical design method using the cavity method.  The cavity method is a method that extends the Bethe approximation to other than two-body interactions.  If the interaction network is a tree graph, the cavity method is rigorous.  The cavity method has been applied to analyze stochastic models on random graphs that can be regarded locally as trees.  In addition, there are a few applications to protein folding problems, such as the computation of phase diagrams for lattice HP models \cite{motanari2004phase} and the prediction of contact maps \cite{weight2009identification}.  However, to the best of our knowledge, there are no studies of protein design using the cavity method.

\section{Model and method}
\subsection{The lattice HP model with interactions between proteins and water}

In the lattice HP model, the backbone chain of a protein is represented by a lattice self-avoiding walk, and each site represents an amino acid residue.  There are only two types of amino acid residues: hydrophobic (H) and hydrophilic (P).  In this study, we consider $N$ residues ${\boldsymbol \sigma} = \{\sigma_{1}, \sigma_{2}, \ldots ,\sigma_{N} |\forall i,\sigma_{i} = 1, 0\}$ on a lattice position $\bm{r} = \{r_{1}, r_{2}, \ldots ,r_{N} \}$, where $i = 1,2, \ldots, N$ $\sigma_{i} = 1$ indicates that the $i$-th residue is an H-residue, and $\sigma_{i} = 0$ indicates that it is a P-residue.  Let $\mu$ be the chemical potential of water, we proposed the following Hamiltonian for proteins with structure $\bm r$ and sequence $\bm \sigma$ in our previous work \cite{takahashi2021lattice}:

\begin{eqnarray}
H(\bm r ,\bm \sigma ; \mu) = - \sum_{i<j} \sigma_{i} \sigma_{j} \Delta(r_{i}-r_{j}) - \mu \sum_{i} (1 - \sigma_{i}).
\end{eqnarray}
The function $\Delta(r_{i} - r_{j})$ in Eq. (1) takes the value 1 only when $r_{i}$ and $r_{j}$ are not continuous in the backbone chain and are nearest neighbors in the coordinate space, otherwise, it tales the values 0. Such a positional relationship is referred to as being in contact with each other.  The first term in Eq. (1) is the Hamiltonian of the lattice HP model usually used.  Because $\sigma_{i}$ is a P-residue, $\sigma_{i} = 0$, Eq. (1) is a modified Hamiltonian of the original Hamiltonian of the lattice HP model by adding the interaction between P-residues and water.

\subsection{Bayesian formalism}
We consider the following probability of a target conformation $\bm r = \bm R$ with a sequence $\bm \sigma$:

\begin{eqnarray}
p(\bm{R} \,|\, \bm{\sigma}) = \frac{1}{Z(\bm \sigma ; \beta, \mu)}e^{-\beta H(\bm{R} , \bm{\sigma} ; \mu)},\\
Z(\bm \sigma ; \beta, \mu) = \sum_{\bm r} e^{-\beta H(\bm{r} , \bm{\sigma} ; \mu)}.
\end{eqnarray}
The partition function (3) is a sum over the states for all possible conformational patterns $\bm r$ that a given sequence can fold.  Henceforth, Eq. (2) is referred to as the target probability or likelihood function. For the target probability (2), the maximum likelihood estimation (in previous studies, using only the first term of Eq. (1)) is the MTP.  However, Eq. (2) contains a partition function (3) for the conformation space, which causes a computational explosion problem, making the MTP extremely difficult in practice.

In our previous work \cite{takahashi2021lattice}, we used the posterior $p(\bm \sigma | \bm R)$ for designing based on Bayesian learning. In the derivation of the posterior $p(\bm \sigma | \bm R)$, we proposed a prior $p(\bm \sigma)$ reflecting the hypothesis explained below.  There, we assumed that evolved sequences have a certain statistical mechanical tendency.  We hypothesize that sequences with smaller free energies have evolved and become more likely to appear.  The prior distribution reflecting this hypothesis is as follows:

\begin{eqnarray}
	p(\bm \sigma) = \frac{Z(\bm \sigma; \beta_{p}, \mu_{p})}{\Xi(\beta_{p}, \mu_{p})}.
\end{eqnarray}
The denominator $\Xi(\beta_{p}, \mu_{p})$ is the partition function given by $\Xi(\beta_{p}, \mu_{p}) = \sum_{\bm \sigma} \sum_{\bm r} e ^ {- \beta_{p} H(\bm{r} ; \bm{\sigma}) }$ and does not depend on conformations and sequences.  Let $\beta_{p}$ and $\mu_{p}$ be the inverse temperature and water chemical potentials in the prior distribution, respectively.  The statistical mechanical explanation of this prior is that the lower the free energy $F(\bm \sigma; \beta_{p}, \mu_{p}) = -(1/ \beta_{p}) \log Z(\bm \sigma; \beta_{p}, \mu_{p})$, the higher the probability (4).

Substituting the target probability (2) and prior (4) into the following Bayes' theorem, we obtain the following posterior:

\begin{eqnarray}
p({\bm \sigma}|\bm R) &= \frac{p(\bm R|\bm \sigma)p(\bm \sigma)}{\sum_{\bm \sigma}p(\bm R|\bm \sigma)p(\bm \sigma)}\\
&\propto \frac{e^{-\beta H(\bm{R} , \bm{\sigma} ; \mu)}}{Z(\bm \sigma ; \beta, \mu)} \cdot \frac{Z(\bm \sigma; \beta_{p}, \mu_{p})}{\Xi(\beta_{p}, \mu_{p})}.
\end{eqnarray}
If $\beta_{p} = \beta$ and $\mu_{p} = \mu$, then the denominator and numerator $Z(\bm \sigma ; \beta, \mu)$ in Eq. (6) cancel each other out.  The partition function $\Xi(\beta_{p}, \mu_{p})$ does not depend on the sequence $\bm \sigma$, so it cancels with the one appearing in the normalization constant in Eq. (5).  Consequently, the following posterior distribution is obtained:

\begin{eqnarray}
	p(\bm \sigma | \bm R) = \frac{e^{-\beta H(\bm{R}, \bm{\sigma} ; \mu) } }{Z(\bm R; \beta, \mu)}, \\
	Z(\bm R; \beta, \mu) = \sum_{\bm \sigma} e^{-\beta H(\bm{R}, \bm{\sigma} ; \mu)}.
\end{eqnarray}
Sampling from the partition function (8) is computationally easier than sampling from the partition function (3).  We perform the Markov chain Monte Carlo (MCMC) method for generating the optimal sequence from the posterior (7) in our previous work \cite{takahashi2021lattice}.

Notably, the derivation of the posterior (7) has interesting theoretical implications.  
This is consistent with the fact that in exact calculations for the internal energy of the spin glass $\pm J$ model, the partition function for the spin configuration cancels out on the Nishimori line \cite{nishimori2002statistical}.  The Nishimori line is the hypersurface in the parameter space \cite{nishimori1980exact} and achieves the Bayes-optimality, an upper bound on accuracy in error-correcting codes \cite{iba1999the}.  Therefore, in other words, our method can be called protein design on the Nishimori line.  In addition, our method provides a correspondence between protein design and error-correcting codes in terms of Bayes-optimality.  Thus, our design theory not only overcomes the computational bottleneck, but also presents a surprising relationship between the ``evolutionary problem", the analytic theory of the spin glass model, and error-correcting code.

\subsection{The cavity method}
Then, we show that posterior (7) can be strictly divided into independent probabilities for each sequence by the cavity method.  This topic is novelty point of this work compared with our previous study.  Posterior (7) can be expressed as follows:

\begin{eqnarray}
p(\bm \sigma | \bm R) = \frac{1}{Z(\bm R; \beta, \mu)} \Bigl(\prod_{a}\psi_{a} (\bm \sigma_{a})\Bigr)
\prod_{i = 1}^{N} \phi_{i}(\sigma_{i}).
\end{eqnarray}
In Eq. (9), let $\bm \sigma_{a}$ be the set of residues related to the $a$-th contact, and $\psi_{a}(\bm \sigma_{a})$ be a function of it. That is, if $\bm \sigma_{a} = \{\sigma_{i}, \sigma_{j}\}$, $\psi_{a} (\bm \sigma_{a}) = \exp(\beta \sigma_{i}\sigma_{j})$. The factor $\phi_{i} (\sigma_{i})$ is defined by $\phi_{i}(\sigma_{i}) = \exp[\beta \mu (1 - \sigma_{i})]$.  The contact graph of the lattice protein is determined by the conformation $\bm R$ explicitly.

We aim to marginalize the posterior (9).  Some isolated residues do not interact with any other residues in a lattice protein.  We illustrate an example the pair of 2D lattice conformation and its contact graph in Fig. 1.  For such isolated residues, marginalization is easy.  The summation of any other residues (not isolated) cancels in the denominator and numerator of Eq. (9); hence the marginal posterior of Eq. (9) is obtained as follows:

\begin{eqnarray}
p(\sigma_{i} | \bm R) = \frac{e^{\beta \mu (1 - \sigma_{i})}}{\sum_{\sigma_{i}} e^{\beta \mu (1 - \sigma_{i})}}.
\end{eqnarray}

\begin{figure}
	\begin{center}
	\vspace{-10mm}
		\includegraphics[width=13cm]{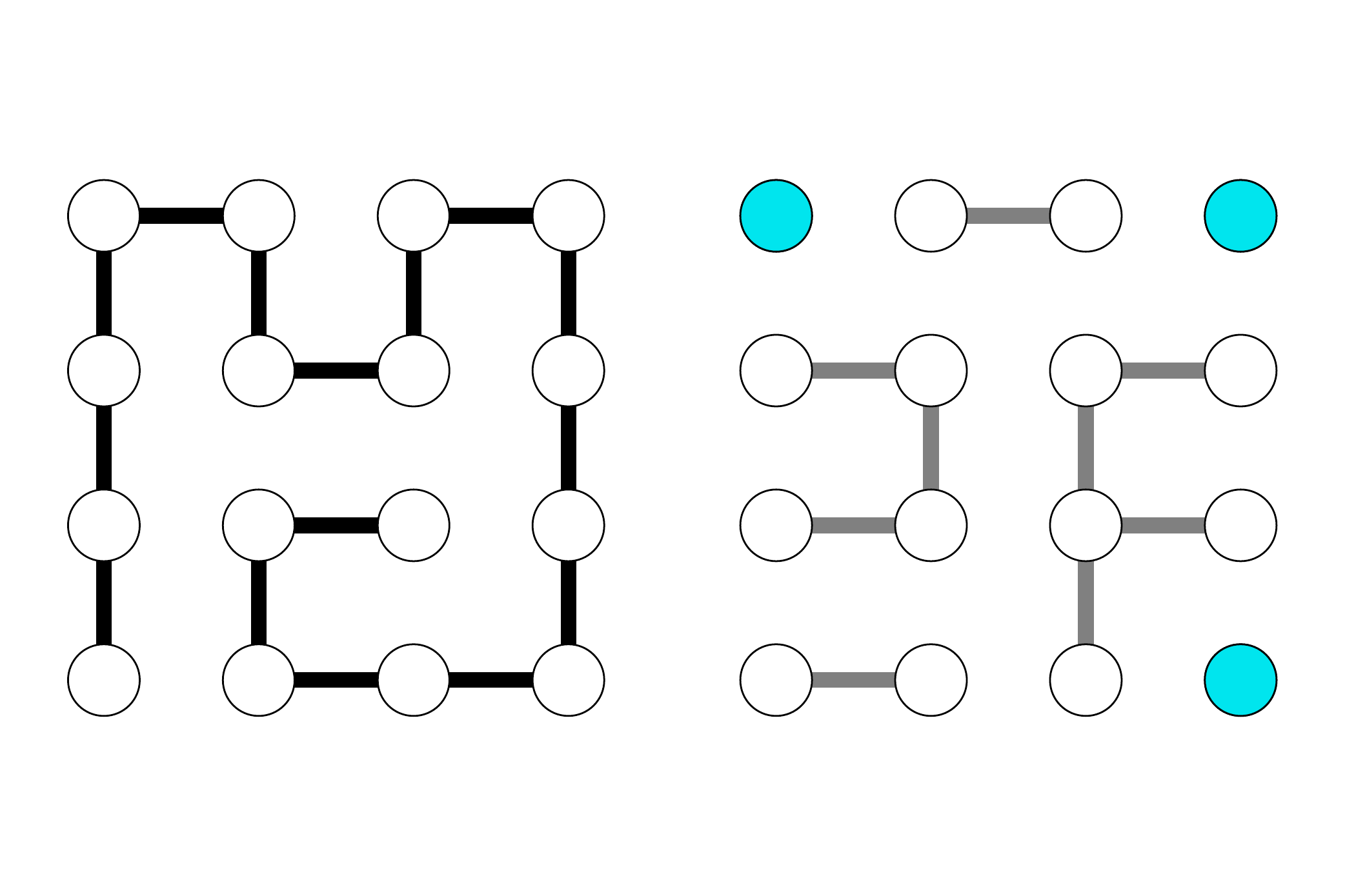}
		\caption{(left): A lattice protein conformation of $N = 4\times4$.  The black line represents the protein backborn structure.   (right): The grey line represents edges of the contact graph of its conformation.  The colored residues are isolated residues.}
	\end{center}	
\end{figure}

For each residue $\sigma_{i}$ in contact with other residues than the isolated one, if the residue-residue interaction network is a tree, one can derive explicitly the recursion formula of the belief propagation (BP), which is an algorithm to compute marginal distributions.   


The BP algorithm is derived by the expectation of Eq. (9) under the probability distribution of the system excluding the residue $\sigma_{i}$: $p_{\setminus i} (\bm \sigma_{\setminus \sigma_{i}} | \bm R)$ (which is called cavity distribution).  We leave the details to Appendix A, only the results are given below:
\begin{eqnarray}
\tilde{\nu}_{a \rightarrow i}^{(t)} (\sigma_{i}) = C_{a \rightarrow i} \sum_{\sigma_{j}} e^{\beta \sigma_{i}\sigma_{j}} \nu_{j \rightarrow a}^{(t)} (\sigma_{j}), \\
\nu_{i \rightarrow a}^{(t + 1)} (\sigma_{i}) = C_{i \rightarrow a} e^{\beta\mu(1 - \sigma_{i})} \prod_{b \in \partial_{i} \setminus a} \tilde{\nu}_{b \rightarrow i}^{(t)} (\sigma_{i}).
\end{eqnarray}
In Eqs. (11) and (12), let $a$ and $b$ be indices on contacts and $i$ and $j$ are indices on residues.  The symbol $\partial_{i}$ denotes the index set of contacts related to residue $\sigma_{i}$. $\tilde{\nu}_{a \rightarrow i}^{(t)} (\sigma_{i})$ is the belief from the $a$-th contact to the $i$-th residue, $\nu_{i \rightarrow a}^{(t + 1)} (\sigma_{i})$ is the belief from the $i$-th residue to the $a$-th contact, and the upper right subscript is the number of steps in the BP algorithm.  The constants $C_{a \rightarrow i}$ and $C_{i \rightarrow a}$ are the normalizing constants of each distribution function.  If one properly defines $\nu_{i \rightarrow a}^{(t = 0)}(\sigma_{i})$ as the initial condition and computes Eqs. (11) and (12) at each step for all combinations $(i, a)$ excluding the isolated residues, after sufficient iterations $t_{\rm max}$, the following belief:

\begin{eqnarray}
	\nu_{i}^{(t)}(\sigma_{i}) = C_{i}\prod_{a \in \partial_{i}} \tilde{\nu}_{a \rightarrow i}^{(t-1)} (\sigma_{i}),
\end{eqnarray}
converges to the marginal distribution $p(\sigma_{i} | \bm R) = \sum_{\bm \sigma \setminus \sigma_{i}} p(\bm \sigma | \bm R)$. In Eq. (13) where $C_{i}$ is the inverse of the normalization constant of $\nu_{i}^{(t)}(\sigma_{i})$, we set the initial condition as a uniform distribution $\nu_{i \rightarrow a}^{(t = 0)}(\sigma_{i} = 1) = \nu_{i \rightarrow a}^{(t = 0)}(\sigma_{i} = 0) = 1/2$.  If $\nu_{i}^{(t)}(\sigma_{i} = 1) > 1/2$ residue $\sigma_{i}$ is H and P otherwise.  Because the above calculations are equivalent to the Ising model of two-body interaction, they are strictly equivalent to the Bethe approximation \cite{kabashima1998belief}.

\subsection{Hyper parameter optimization}

The optimal hyperparameter, the chemical potential of water $\mu = \mu^{*}$, is the same value obtained in our previous study \cite{takahashi2021lattice}.  The optimal chemical potential $\mu^{*}$ was determined to be the value with the highest design accuracy by repeated computational experiments.

This procedure does not mean that our design method needs to carry out MCMC for determination of $\mu^{*}$ before using the cavity method.  Of course, the proposed design method using the cavity method also can determine $\mu^{*}$ in the same way.  In this study, as a primary goal, we investigate whether the cavity method can achieve the same design accuracy as MCMC or not in the same conditions.  Thus, we use the same value of $\mu^{*}$ obtained by MCMC.

The criterion of optimization of the hyper parameter $\mu$ is nontrivial.  For instance, the minimization of the Bethe free energy, corresponding to the maximization of the marginal likelihood in the context of Bayesian learning, is one of promising candidates for the criterion. However, whether or not the maximization of the marginal likelihood maximizes the design accuracy is not clear.  In this proposed design method, the design criterion is the maximizer of posterior marginals (MPM) according to the procedure described in the previous section. The hyper parameter $\mu$ was determined to be the value with the highest design accuracy.  Thus, in this study, the criterion of optimization of the hyper parameter $\mu$ is the maximization of the design accuracy under the MPM.

\section{Results}
\subsection{2D small conformations}

We now show the design results.  In this study, we use 2D small lattice proteins for which all possible conformations are enumerable, allowing us to determine whether the design was successful or not rigorously because one can calculate the energies of all pairs of the generated sequence and every conformation.  Specifically, we use the lattice conformations of $N = 9, 12$ and 16 involving not maximally compact conformations used in the previous study \cite{irback2002enumerating}.  The reason of why we use not only square lattice but also non square lattice is that the native conformations are not necessarily maximally compact. This is because proteins can have low energy if the hydrophobic core is compact enough \cite{yue1995forces}. Moreover, we design more large size lattice proteins, $5\times5$, and $6\times6$ square lattices.  Because $N = 5\times5$ and $6\times6$ have many total conformations (1,075 and 52,667, respectively), we chose 100 conformations randomly for each size due to computational cost.  Also, we do not design non-square conformations of $N = 5\times5$ and $6\times6$ because the number of patters of such lattices is too many to generate. The meaning of "designable" here is the number of sequences that make a target conformation the only ground state (designability) is nonzero.

We did not use three-dimensional (3D) lattice proteins.  The reason is that the structures $N = 2\times2\times3$ and $N = 3\times3\times3$, for which one can determine strict design success, are not typical examples of proteins because the number of core residues relative to the number of surface residues is deficient compared with natural proteins.

There are three types of sequences: good sequence, which has the target conformation as a unique ground state; medium sequence, which has the target conformation as one of the degenerated ground state; bad sequence, which has the ground state conformation(s) that does not include the target conformation. 

We summarize our design results in Table I.  Table I shows the sequences generated using the cavity method and our previous work by MCMC, where $N_{\rm c}^{(\rm g)}$, $N_{\rm c}^{(\rm m)}$, and $N_{\rm c}^{(\rm b)}$ are the number of structures that successfully obtained a good, medium, and bad sequences, respectively.  Therefore, the design success rate, SR is the ratio of $N_{\rm c}^{(\rm g)}$ to the total number of structures $N_{\rm c}$. The optimal value of chemical potential $\mu^{*}$ was determined as explained above, and the same value was used for the cavity method and MCMC.  The value $\mu^{*}$ may differ for each conformation even if the size is the same, but we use the same $\mu^{*}$ for the same size without considering this issue.  The total number of conformations designed is the same for the cavity method and MCMC. 

\begin{table}[]
	\begin{center}
		\caption{Comparison of the cavity method and MCMC design results. The hyperparameter $\mu^{*}$ was calculated many times for each size in MCMC case, and the values achieve the highest success rate, SR.  The same values of $\mu^{*}$ were used for the cavity method.}
			\begin{tabular}{@{\extracolsep{\fill}} ccccc|ccccc}\hline \hline
			& \multicolumn{4}{c|}{Cavity method} &  \multicolumn{4}{c}{MCMC}\\
			Size&$N_{\rm c}^{(\rm g)}$ & $N_{\rm c}^{(\rm m)}$ & $N_{\rm c}^{(\rm b)}$ & SR (\%) & $N_{\rm c}^{(\rm g)}$ & $N_{\rm c}^{(\rm m)}$ & $N_{\rm c}^{(\rm b)}$ & SR (\%) & $\mu^{*}$ \\ \cline{1-2}\hline
			\hspace{0.5mm}$N = 9$ & 7 & 1 & 0 & 87.5& 7 & 1 & 0& 87.5 &0.55 \\
			\hspace{0.5mm}$N = 12$ & 29 & 11 & 0 & 72.5 & 29 & 11 & 0 &72.5 & 0.6 \\
			\hspace{0.5mm}$N = 16$ & 393 &89 &0 & 81.5 & 393 & 89 & 0 & 81.5 &0.62\\
			\hspace{0.5mm}$5\times5$ & 68 & 32 & 0 & 68 & 68 & 32 & 0 & 68 &0.74\\
			\hspace{0.5mm}$6\times6$ & 62 &38&0& 62 & 63 & 37 & 0 & 63 &0.8\\ 
			\hline \hline		
			\end{tabular}
			\label{tab:Compact lattice models and design results}
	\end{center}
\end{table}

Table I shows that the cavity method and MCMC differ slightly in the percentage of correct answers at $6\times6$, otherwise, they perform precisely the same. This result shows almost no difference between the cavity method and MCMC in design accuracy, at least for small 2D lattice proteins.  The conformations for the $N_{\rm c}^{(\rm g)}$, $N_{\rm c}^{(\rm m)}$, and $N_{\rm c}^{(\rm b)}$ are same in the cavity method and MCMC other than the case of $N = 6\times6$.  In the case of $N = 6\times6$, one conformation is designed successfully by the cavity method but MCMC failed to design it, two conformations are designed successfully by MCMC but the cavity method failed to design those two.

\subsection{Large 2D conformations}

Here, we show the result for 2D lattice proteins larger than in the previous subsection to test the design method using the cavity method for more realistic protein models.  We chose two 2D lattice proteins with comparatively large size ($N = 23, 50$) models designed by the MTP-based method of Irb\"{a}ck {\it et al.} \cite{irback1998monte, irback1999design}.  That confirmed that the designed sequence would likely fold into the target conformation with simulated tempering.  We already designed those two conformations in the previous study using MCMC and succeeded to design these two conformations with the same sequences \cite{takahashi2021lattice}.  Therefore, we use the same values of the optimal chemical potentials of the previous study: $\mu^{*} = 0.7$  (0.85)  for $N = 23$ (50), respectively.

As a result, using the cavity method, we designed those two with the same sequence of \cite{irback1998monte, irback1999design}.  Fig. 2 shows those two designed conformations.   The white balls represent H-residues, and the black balls represent P-residues.

\subsection{Comparison of the calculation time}

We also compare the calculation time of the cavity method and MCMC for the $N = 50$ (right-hand side of Fig. 2) using a standard PC (Apple M1 MacBook Pro with 8 GB memory).  As a result, the calculation time of both cases is 5.282 seconds by MCMC and 1.433 seconds by the cavity method (BP).  In the case of BP, one can carry out the identical calculation for each residue.  We, therefore, carried out the parallel computation by four threads for BP, and the result was 0.506 seconds.  The cavity method is about three times faster than MCMC with no parallelization and is about ten times faster with parallelization (four threads).  Parallel computation is one of the significant advantages of the cavity method.

	\begin{figure}
		\begin{center}
		\vspace{-10mm}
			\includegraphics[width=13cm]{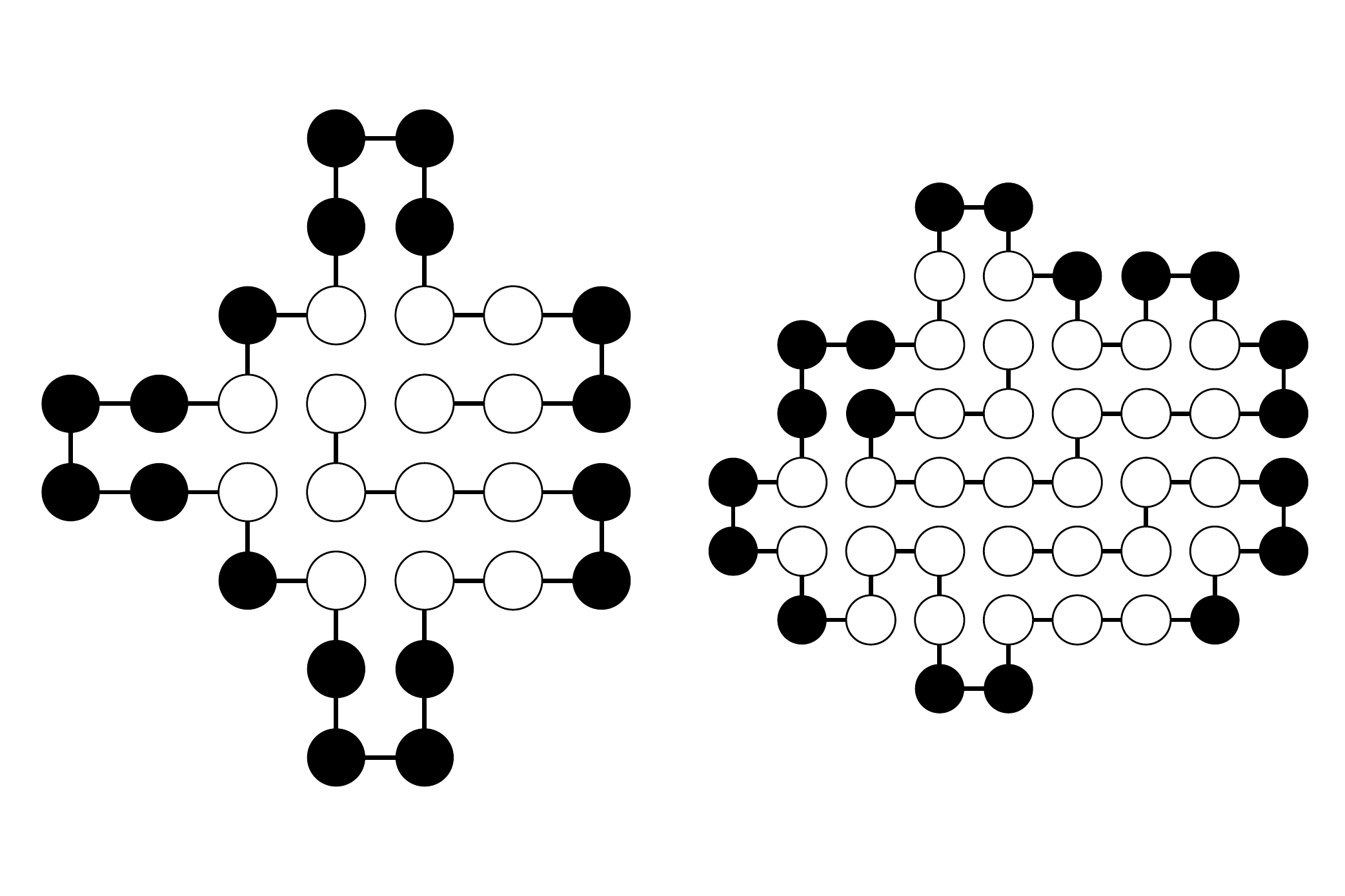}
			\caption{(left): Designed conformation of $N = 23$ with $\beta = 10$ and $\mu^{*} = 0.7$. (right): Designed conformation of $N = 50$ with $\beta = 10$ and $\mu^{*} = 0.85$.  The white balls represent H-residues, and the black balls represent P-residues.  Our present design method succeeded to design these two with the identical sequences of conventional studies \cite{irback1998monte, irback1999design, takahashi2021lattice}.  Our current approach using the cavity method can be most efficient because it skips internal conformational search due to the partition function (3), and it uses the recursion of BP algorithm instead of MC sampling.}
		\end{center}	
	\end{figure}

\section{Discussion}

\subsection{Contact graph of proteins}

What does it mean that the results from the cavity method, a generalization of the Bethe approximation, and the numerical results from MCMC are almost identical?  That is, simply put, the lattice protein contact graph has a graph structure suited to the Bethe approximation.  In addition, because the cavity method can compute the posterior probabilities around each amino acid type independently for each residue, the parallel computation beneficially reduces the computation time with an increase in protein size. 

It is unclear whether the contact graph of any real protein is a tree.  In addition, although the loop effect can be ignored in the thermodynamic limit, it is unclear whether the loop effect can be ignored for the contact graphs of real proteins with a finite number of amino acid residues (hundreds to thousands).  However, because our approach is equivalent to the Bethe approximation, we can expect it to be a good approximation if the fluctuations in the mean-field of the next-nearest neighbor residues are sufficiently small.  Therefore, we believe that extending our approach to real proteins is worthwhile.  We are currently verifying this using data on the 3D structure and amino acid sequence of real proteins and will report on our findings soon.

\subsection{Biological evidence of the prior}

The prior Eq. (4) implies that sequences enriched in polar residues are likely to be evolutionarily selected for relatively large $\beta$ and $\mu$ values, owing to the effect of the second term in Eq. (1). This implication is consistent with the fact that organisms have many intrinsically disordered proteins (IDPs), which are characterized by a high proportion of polar residues and lack of an ordered three-dimensional structure; for example, Oates et al. estimated that the percentage of disordered residues in the human proteome is between 37\% and 50\% \cite{oates2013database}. IDPs are important components of the cellular signaling machinery, allowing the same polypeptide to undertake different interactions with different partners \cite{wright2015intrinsically}. In addition, recent studies have shown that IDPs play an important role in formation of membraneless organelles, enabling internal spatiotemporal control of complex biochemical reactions in a cell \cite{boeynaems2018protein}. These observations suggest that the physical property of the prior Eq. (4) is advantageous for cells to efficiently perform complex chemical reactions.

\addcontentsline{toc}{section}{Acknowledgment}
\section*{Acknowledgment}

This work was financially supported by JST SPRING, Grant Number JPMJSP2125. The author (Initial) would like to take this opportunity to thank the ``Interdisciplinary Frontier Next-Generation Researcher Program of the Tokai Higher Education and Research System".  The authors are grateful to T. Obuchi, Kyoto University for illuminating discussions.  This work was also supported by KAKENHI Nos. 19H03166 (G.C.) and 19K03650 (K.T.).

\addcontentsline{toc}{section}{Appendix A}
\section*{Appendix A.  Derivation of the update rules of belief propagation for lattice proteins}
\renewcommand{\theequation}{A.\arabic{equation}}
\setcounter{equation}{0}

We first consider the cavity distribution of the posterior (9) in the main text.  Cavity distribution is the joint probability distribution of the system without residues which do not relate to $\sigma_{i}$. Its formula is given by

\begin{eqnarray}
	p_{\setminus i} (\bm \sigma_{\setminus \sigma_{i}} | \bm R) = \frac{\prod_{b \notin \partial_{i}} \psi_{b} (\bm \sigma_{b}) \prod_{j \neq i} \phi_{j}(\sigma_{j})}{\sum_{\bm \sigma_{\setminus \sigma_{i}}} \prod_{b \notin \partial_{i}} \psi_{b} (\bm \sigma_{b}) \prod_{j \neq i} \phi_{j}(\sigma_{j})},
\end{eqnarray}
where $\psi_{b} (\bm \sigma_{b})$ is given by $\psi_{b} (\bm \sigma_{b}) = \exp(\beta \sigma_{j}\sigma_{k})$ if $\bm \sigma_{b} = \{\sigma_{j}, \sigma_{k}\}$. The factor $\phi_{j}(\sigma_{j})$ is defined by $\phi_{j}(\sigma_{j}) = \exp[\beta \mu (1 - \sigma_{j})]$.  Then, for all systems, the marginal distribution $p(\sigma_{i} | \bm R) = \sum_{\bm \sigma_{\setminus \sigma_{i}}} p(\bm \sigma | \bm R)$ can be rigorously expressed using the cavity distribution (A.1) as follows:

\begin{eqnarray}
	p(\sigma_{i} | \bm R) = \frac{\braket{\prod_{a \in \partial_{i}} \psi_{a} (\bm \sigma_{a}) \phi_{i}(\sigma_{i})}_{\setminus \sigma_{i}}}{\sum_{\bm \sigma_{\setminus \sigma_{i}}} \braket{\prod_{a \in \partial_{i}} \psi_{a} (\bm \sigma_{a}) \phi_{i}(\sigma_{i})}_{\setminus \sigma_{i}}},
\end{eqnarray}
where $\braket{\cdot}_{\setminus \sigma_{i}}$ means the expectation by the cavity distribution (A.1).

\begin{proof}
	We separate the posterior (9) in the main text into the part that includes a and the part that does not include $\sigma_{i}$,

	\begin{eqnarray}
	\fl	p(\bm \sigma | \bm R) = \frac{
		\Bigl(\prod_{a \in \partial_{i}} \psi_{a}(\bm \sigma_{a}) \prod_{i \neq j}\phi_{i}(\sigma_{i})\Bigr)
		\Bigl(\prod_{b \notin \partial_{i}} \psi_{b}(\bm \sigma_{b}) \prod_{j \neq i} \phi_{j}(\sigma_{j})\Bigr)}
		{\sum_{\bm \sigma} \Bigl(\prod_{a \in \partial_{i}} \psi_{a}(\bm \sigma_{a}) \prod_{i \neq j}\phi_{i}(\sigma_{i})\Bigr)
		\Bigl(\prod_{b \notin \partial_{i}} \psi_{b}(\bm \sigma_{b}) \prod_{j \neq i} \phi_{j}(\sigma_{j})\Bigr)}.
	\end{eqnarray}
	We divide the denominator and the numerator of right hand side of Eq. (A.3) by the constant $\sum_{\bm \sigma \setminus_{\sigma_{i}}} \prod_{b \notin \partial_{i}} \psi_{b}(\bm \sigma_{b}) \prod_{j \neq i} \phi_{j}(\sigma_{j})$.  The numerator of (A.3) is then
	\begin{eqnarray*}
		 \frac{\Bigl(\prod_{a \in \partial_{i}} \psi_{a}(\bm \sigma_{a}) \prod_{i \neq j}\phi_{i}(\sigma_{i})\Bigr)
		\Bigl(\prod_{b \notin \partial_{i}} \psi_{b}(\bm \sigma_{b}) \prod_{j \neq i} \phi_{j}(\sigma_{j})\Bigr)}{\sum_{\bm \sigma \setminus_{\sigma_{i}}} \prod_{b \notin \partial_{i}} \psi_{b}(\bm \sigma_{b}) \prod_{j \neq i} \phi_{j}(\sigma_{j})} \\
		= 
		\prod_{a \in \partial_{i}} \psi_{a}(\bm \sigma_{a}) \prod_{i \neq j}\phi_{i}(\sigma_{i}) p_{\setminus i} (\bm \sigma_{\setminus \sigma_{i}} | \bm R), 		
	\end{eqnarray*}
			and the dominator is calculated similarly.  Hence,	
	\begin{eqnarray*}
		p(\bm \sigma|\bm R) = \frac{\prod_{a \in \partial_{i}} \psi_{a}(\bm \sigma_{a}) \prod_{i \neq j}\phi_{i}(\sigma_{i}) p_{\setminus i} (\bm \sigma_{\setminus \sigma_{i}} | \bm R)}{\sum_{\bm \sigma} \prod_{a \in \partial_{i}} \psi_{a}(\bm \sigma_{a}) \prod_{i \neq j}\phi_{i}(\sigma_{i}) p_{\setminus i} (\bm \sigma_{\setminus \sigma_{i}} | \bm R)}.
	\end{eqnarray*}
	Finally, by the marginalization $p(\sigma_{i}| \bm R) = \sum_{\bm \sigma \setminus_{\sigma_{i}}} p(\bm \sigma|\bm R)$, we obtain
	\begin{eqnarray*}
		p(\sigma_{i}| \bm R) &=& \frac{\sum_{\bm \sigma \setminus_{\sigma_{i}}} \prod_{a \in \partial_{i}} \psi_{a}(\bm \sigma_{a}) \prod_{i \neq j}\phi_{i}(\sigma_{i}) p_{\setminus i} (\bm \sigma_{\setminus \sigma_{i}} | \bm R)}
		{\sum_{\bm \sigma} \prod_{a \in \partial_{i}} \psi_{a}(\bm \sigma_{a}) \prod_{i \neq j}\phi_{i}(\sigma_{i}) p_{\setminus i} (\bm \sigma_{\setminus \sigma_{i}} | \bm R)}\\
		&=& \frac{\sum_{\bm \sigma \setminus_{\sigma_{i}}} \prod_{a \in \partial_{i}} \psi_{a}(\bm \sigma_{a}) \prod_{i \neq j}\phi_{i}(\sigma_{i}) p_{\setminus i} (\bm \sigma_{\setminus \sigma_{i}} | \bm R)}
		{\sum_{\sigma_{i}} \sum_{\bm \sigma \setminus_{\sigma_{i}}} \prod_{a \in \partial_{i}} \psi_{a}(\bm \sigma_{a}) \prod_{i \neq j}\phi_{i}(\sigma_{i}) p_{\setminus i} (\bm \sigma_{\setminus \sigma_{i}} | \bm R)}\\
		&=& \frac{\braket{\prod_{a \in \partial_{i}} \psi_{a} (\bm \sigma_{a}) \phi_{i}(\sigma_{i})}_{\setminus \sigma_{i}}}{\sum_{\bm \sigma_{\setminus \sigma_{i}}} \braket{\prod_{a \in \partial_{i}} \psi_{a} (\bm \sigma_{a}) \phi_{i}(\sigma_{i})}_{\setminus \sigma_{i}}}.
	\end{eqnarray*}
\end{proof}

We define following effective potential $\psi_{i}^{\rm eff}(\sigma_{i})$ by the numerator of right hand side of Eq. (A.2):

\begin{eqnarray}
	\psi_{i}^{\rm eff}(\sigma_{i}) = \sum_{\bm \sigma \setminus_{\sigma_{i}}} \Bigl(\prod_{a \in \partial_{i}} \psi_{a}(\bm \sigma_{a}) \Bigr) \phi_{i}(\sigma_{i}) p_{\setminus i} (\bm \sigma_{\setminus \sigma_{i}} | \bm R).
\end{eqnarray}
Additionally, for $a \in \partial_{i}$, we consider the system excluding $\psi_{a}(\bm \sigma_{a})$, called a-cavity system.  We define the marginal distribution of $\sigma_{j}$ which is included in $\bm \sigma_{a}$ under $a$-cavity system by $\nu_{j \rightarrow a}(\sigma_{j})$.  If the contact graph of given target conformation is a tree,  excluding any residue $\sigma_{i}$ makes the contact graph divided into independent part per residues related to the contact $a$.

Therefore, if the contact graph of given target conformation is a tree, we can calculate (A.4) as follows:

\begin{eqnarray}
	\psi_{i}^{\rm eff}(\sigma_{i}) &=& \phi_{i}(\sigma_{i}) \prod_{a \in \partial_{i}} \Bigl(\sum_{\bm \sigma_{a} \setminus_{\sigma_{i}}} \psi_{a}(\bm \sigma_{a}) 
	\sum_{\bm \sigma \setminus_{\bm \sigma_{a}}}  p_{\setminus i} (\bm \sigma_{\setminus \sigma_{i}} | \bm R) \Bigr) \nonumber \\
	&=&  \phi_{i}(\sigma_{i}) \prod_{a \in \partial_{i}} \Bigl(\sum_{\bm \sigma_{a} \setminus_{\sigma_{i}}} \psi_{a}(\bm \sigma_{a})\prod_{j \in \partial_{a} \setminus_{i}} \nu_{j \rightarrow a}(\sigma_{j})
	 \Bigr) \nonumber \\
		&=& \phi_{i}(\sigma_{i}) \prod_{a \in \partial_{i}} \Bigl(\sum_{\sigma_{j}} \psi_{a}(\bm \sigma_{a}) \nu_{j \rightarrow a}(\sigma_{j}) \Bigr). \hspace{15mm}(j \in \partial_{a} \setminus_{i})
\end{eqnarray}
In Eq. (A.5), we used $\sum_{\bm \sigma_{a} \setminus_{\sigma_{i}}} = \sum_{\sigma_{j}}$ and $\prod_{j \in \partial_{a} \setminus_{i}} \nu_{j \rightarrow a}(\sigma_{j}) = \nu_{j \rightarrow a}(\sigma_{j})$, because each index set $\partial_{a}$ has only two indices in the lattice HP model. 

Then, we consider the effective potential of $a$-cavity system $\psi_{i \rightarrow a}^{\rm eff} (\sigma_{i})$.  $\psi_{i \rightarrow a}^{\rm eff} (\sigma_{i})$ is obtained by excluding $\psi_{a}(\bm \sigma_{a})$ from $\psi_{i}^{\rm eff}(\sigma_{i})$.  Hence we obtain $\psi_{i \rightarrow a}^{\rm eff} (\sigma_{i})$ as follows:

\begin{eqnarray*}
	\psi_{i \rightarrow a}^{\rm eff} (\sigma_{i}) = \phi_{i}(\sigma_{i}) \prod_{b \in \partial_{i} \setminus_{a}} \Bigl(\sum_{\sigma_{k}} \psi_{b}(\bm \sigma_{b}) \nu_{k \rightarrow b}(\sigma_{k}) \Bigr).\hspace{10mm} (k \in \partial_{b} \setminus_{i})
\end{eqnarray*}
From the definition of $\nu_{j \rightarrow a}(\sigma_{i})$ explained above, $\nu_{j \rightarrow a}(\sigma_{i})$ is obtained by normalization $\psi_{i \rightarrow a}^{\rm eff} (\sigma_{i})$.  Therefore, let $j = k$, we obtain following expression:

\begin{eqnarray*}
	\nu_{j \rightarrow a}(\sigma_{i}) = \frac{\phi_{i}(\sigma_{i}) \prod_{b \in \partial_{i} \setminus_{a}} \Bigl(\sum_{\sigma_{j}} \psi_{b}(\bm \sigma_{b}) \nu_{j \rightarrow b}(\sigma_{j}) \Bigr)}{\sum_{\sigma_{i}}\phi_{i}(\sigma_{i}) \prod_{b \in \partial_{i} \setminus_{a}} \Bigl(\sum_{\sigma_{j}} \psi_{b}(\bm \sigma_{b}) \nu_{j \rightarrow b}(\sigma_{j}) \Bigr)}. \hspace{10mm} (j \in \partial_{b} \setminus_{i})
\end{eqnarray*}

Let $\tilde{\nu}_{a \rightarrow i}(\sigma_{i})$ be the distribution function derived by normalization $\sum_{\sigma_{j}} \psi_{a}(\bm \sigma_{a}) \nu_{j \rightarrow a}(\sigma_{j})$, the ``belief'' from $a$ to $\sigma_{i}$.  Then, we obtain following expression of two distribution functions:
\begin{eqnarray*}
	\tilde{\nu}_{a \rightarrow i} (\sigma_{i}) &=& C_{a \rightarrow i} \sum_{\sigma_{j}} \psi_{a}(\bm \sigma_{a}) \nu_{j \rightarrow a} (\sigma_{j}), \hspace{10mm} (j \in \partial_{b} \setminus_{i})\\
\nu_{i \rightarrow a} (\sigma_{i}) &=& C_{i \rightarrow a} \phi_{i}(\sigma_{i}) \prod_{b \in \partial_{i} \setminus a} \tilde{\nu}_{b \rightarrow i} (\sigma_{i}),
\end{eqnarray*}
where $C_{a \rightarrow i}$ and $C_{i \rightarrow a}$ are normalization constant of each distribution functions, respectively.  When $\partial_{a} = \{i,j\}$, one can use $\psi_{a}(\bm \sigma_{a}) = e^{\beta\sigma_{i}\sigma_{j}}$.  In addition, using $\phi_{i}(\sigma_{i}) = \exp[\beta \mu (1 - \sigma_{i})]$, we obtain the update rules of BP: Eq. (11) and Eq. (12) in the main text.

\addcontentsline{toc}{section}{References}
\section*{References}
\bibliography{Takahashi_JSTAT_resubmit_2022_8_26.bib}
\bibliographystyle{unsrt}

\end{document}